\documentclass[conference]{IEEEtran}
\IEEEoverridecommandlockouts
\usepackage{amsmath,amsfonts}
\usepackage{array}
\usepackage[caption=false,font=normalsize,labelfont=sf,textfont=sf]{subfig}
\usepackage{textcomp}
\usepackage{stfloats}
\usepackage{url}
\usepackage{verbatim}
\usepackage{graphicx}
\usepackage{cite}
\usepackage{url}
\usepackage{ulem}
\usepackage{lineno,hyperref}
\usepackage{amsmath,amssymb, amsfonts}
\usepackage{graphicx}
\usepackage{textcomp}
\usepackage{xcolor}
\usepackage{array}
\usepackage{url}
\usepackage{lineno,hyperref}
\usepackage{color,framed}
\usepackage{color,xcolor}
\usepackage{algorithm}  
\usepackage{threeparttable}
\usepackage{multirow}
\usepackage{longtable}
\usepackage{ulem}
\usepackage{array}
\usepackage{booktabs}
\usepackage{amsmath}
\usepackage{amssymb}
\usepackage{lineno,hyperref}
\usepackage{listings,xcolor}
\usepackage{courier}
\usepackage{lineno}
\usepackage[justification=centering]{caption}
\usepackage{pdflscape}
\usepackage{bbding}
\usepackage{adjustbox}
\usepackage{enumerate}
\usepackage{blkarray}

\newcommand{\reffig}[1]{Fig.\ref{#1}}

\def\BibTeX{{\rm B\kern-.05em{\sc i\kern-.025em b}\kern-.08em
T\kern-.1667em\lower.7ex\hbox{E}\kern-.125emX}}
\begin{document}

\title{Challenges of Blockchain adoption in financial services in China's Greater Bay Area\\
}

\author{
  \IEEEauthorblockN{Xiongfei Zhao,}
  \IEEEauthorblockA{Department of Computer and Information Science\\
    University of Macau, Macau\\
    \textit{\href{mailto:yb97480@um.edu.mo}{yb97480@um.edu.mo}}\\
  }
  \and
  \IEEEauthorblockN{Yain-Whar Si,}
  \IEEEauthorblockA{Department of Computer and Information Science\\
    University of Macau, Macau\\
    \textit{\href{mailto:fstasp@@um.edu.mo}{fstasp@@um.edu.mo}}\\
  }%
}

\maketitle

\begin{abstract}
In China's Greater Bay Area (Guangdong-Hong Kong-Macao), the increasing use of Blockchain technology in financial services has the potential to generate benefits for many stakeholders. Blockchains are known for their distinctive features, such as decentralized architecture, tamper-proof data structures, and traceable transactions. These features make Blockchain a preferred choice of platform for developing applications in financial service areas. Meanwhile, some questions have been raised regarding Blockchain's suitability to compete with or even replace existing financial systems. This paper provides insights into the current progress of Blockchain applications in insurance, banking, payments, asset trading, loans, remittances, the Internet of Things (IoT) for the finance industry, financial inclusions, and enterprise-level interaction in finance and governance. We review the barriers to widespread Blockchain adoption, especially the risks when transaction fees dominate mining rewards. By comparing the emerging Blockchain technologies and incentive issues related to real-world applications, we hope that this paper can serve as a valuable source of reference for Blockchain researchers and developers in financial service areas.
\end{abstract}

\vspace{-0.2cm}
\section{Introduction}

\textcolor{black}{The Guangdong, Hong Kong, Macao Greater Bay Area (GBA) is a region of great significance, encompassing Hong Kong, Macao, Guangzhou, and Shenzhen as the core engines of regional development. The GBA boasts a unique development structure that encompasses two political systems, three customs zones, three currencies, and different legal systems. Currently, several Blockchain platforms have been adopted in the Greater Bay Area, including banking, assets certification, health information management, and production management of traditional Chinese medicine, etc. Although the financial industry is actively adopting Blockchain technology in their business operations, integrating financial resources in GBA to break down the circulation barriers, regulatory barriers, and talent barriers remains a significant challenge for the region's development.}

Blockchain is a distributed ledger that records all transaction information securely. The essence of this technology is that different nodes participate in a distributed database, and each node has all the data in the network. Transactions are packed into a block that contains a pointer to a previous block. As a result, a Blockchain is formed by linking the transaction information of different periods. Through these features, Blockchain technology becomes one of the main drivers behind digital currencies and a potential foundation for value transfer and public ledger. 



Numerous financial institutions are actively embracing and striving to become pioneers in their respective fields by participating in the development and application of Blockchain technology. Financial institutions, in particular, are showing a preference for Private Blockchain or Consortium Blockchain that do not involve mining incentives, as they allow them to focus on transforming their businesses to the new network while maintaining their existing profit models. However, the need for validators to act honestly is critical, and this can lead to unnecessarily strict and inefficient verification of participants' qualifications. Despite this, the successful implementation of Blockchain in financial services areas is not as prevalent as anticipated. In light of this phenomenon, we present the following insights:


\textbf{\textsl{Obstacles to widespread adoption of Blockchain.}} After migrating the traditional financial services to the Blockchain platform, the related revenue is still tied to the existing model. Meanwhile, as participants of Blockchain, there are differences in business scope, volume, and fee-charging standards among different financial institutions. As a result, the incentive model for participating in the Blockchain infrastructure is not always attractive to all the participants.

\textbf{\textsl{Threats to using transaction fees as block rewards.}} As a typical example of financial services, a Blockchain-based asset trading system requires investors to pledge transaction fees as mining rewards. However, when transaction fee dominates mining rewards, strategic deviations such as \textsl{Selfish Mining}, \textsl{Undercutting}, \textsl{Mining Gap}, \textsl{Pool Hopping} \cite{10.1145/2976749.2978408} could threaten the integrity and security of the Blockchain.

\textbf{\textsl{Blockchain selection for financial services.}} Blockchain as a financial infrastructure requires a variety of qualities to facilitate rich financial scenarios. We compare the current mainstream Blockchain platforms in terms of network type, support of non-fungible tokens (NFT), support of cryptocurrency, enabling smart contracts, consensus type, mining costs, and throughput. \textcolor{black}{From the comparison results, we conclude that Blockchain such as Electro-Optical System (EOS) is more suitable to be adopted as a reference for building the financial infrastructure due to its superior features.}

It is worth noting that public Blockchain is valuable and crucial for fostering healthy competition and collaboration among financial institutions. If the transaction fee is used as the mining reward, the Blockchain-based model is more in line with the current profit model of financial businesses, making it more sustainable than issuing tokens to maintain competition among miners. The insights presented in this paper can be extrapolated to other application areas that span multiple jurisdictions or countries.

This paper is organized as follows. Section 2 reviews the recent research status of Blockchain and introduces the typical financial applications of Blockchain. Section 3 describes the obstacles to widespread Blockchain adoption in financial service areas. Section 4 further explains the threats of using transaction fees as an incentive for mining in order to facilitate the adoption. In section 5, we compared and evaluated the most suitable Blockchain technologies for financial adoption in GBA. The paper is concluded in Section 6.

\section{Blockchain applications in financial service areas}

Blockchain technology has emerged as a game-changing innovation in the financial service industry, offering secure, transparent, and efficient solutions to various challenges faced by traditional financial systems. This technology has the potential to transform the way financial services operate in areas such as insurance, banking, payment, asset trading, loan, and remittance. In this article, we will explore the various ways in which Blockchain technology can be applied to these financial service areas.

\subsection{Insurance}

Existing insurance systems have already archived a certain degree of capability in automatic processing as well as in premium settlement. However, current technologies are still inadequate, especially for managing processes between multiple transacting parties. In order to achieve consistency between businesses and financial data, manual intervention is often required for the reconciliation at the business and financial levels. In certain situations, the customer's experience is affected because payment information cannot be provided to the customers as well as to the providers immediately after the purchase of insurance products. Such incoherence could be eliminated by processing insurance on Blockchain-enabled platforms. Blockchain-based processing systems may adopt smart contracts to manage transactions of insurance products to improve the efficiency of insurance processing. Blockchain-enabled solutions could also eliminate insurance fraud, ensure data security, and reduce operation and management costs. 

In \cite{8967468}, Veneta et al. present an experimental implementation of insurance services based on Ethereum smart contracts. This solution can either be implemented in a public Blockchain, a private Blockchain, or a combined solution in which insurance claims are automatically processed on the private Blockchain, and automatic payments are realized on the public Blockchain. Meskini and Aboulaich \cite{8942270} proposed a Takaful-principles-based insurance model which is based on Consortium Blockchain with smart contracts. They use a complex system simulation software called ``Netlogo'' to study customers' preferences and behavior changes under multi-agent iterations. The result indicates that the transparency and effectiveness of the Blockchain-enabled insurance system have advantages in influencing and attracting customers.

\subsection{Banking}

Banking is based on trust, meaning that banks are entrusted to provide clients with reliable transaction processing, asset management, and investment intermediaries. More importantly, banks can ensure that customers' personal information, asset information, transaction records, investment portfolios, and other private information are not disclosed without authorization. To meet these challenges, traditional banking systems need to manage large volumes of transactions and safeguard generated data. 

Although Blockchains are being considered for the banking industry, there is a high risk of information leakage due to the distributed storage characteristics. Therefore, only a few banks are adopting Blockchain technology, and there are still challenges that remain for the banking industry worldwide. For customer data protection, Ma et al. \cite{8636292} proposed a customer data disclosure confirmation scheme through Nudge Theory to establish the data privacy protection framework for Blockchain. The proposed scheme was based on the predefined data privacy classification and collaborative-filtering-based model. For transaction data protection, Popova N. A. and Butakova N. G \cite{8657279} suggested using Blockchain technology without tokens to protect transaction information such as name, card number, amount, etc.

\subsection{Payment}

Blockchain is considered to be an ideal carrier for payment systems due to its robustness, credibility, transparency, and non-reproducibility properties. Despite these useful features, one of the most prominent problems is how to improve the scalability of Blockchain-based payment systems. By relying on a synchronized global state, current Blockchain systems suffer from extremely low throughput and high latency. For example, with a transaction processing efficiency of seven transactions per second, Bitcoin can validate every transaction and store them forever. In contrast, credit card companies currently process more than 10,000 transactions per second which is more than 1,000 times the throughput of Bitcoin transactions. Recently, two new approaches for improving Blockchain scalability were proposed: 

The first approach is to use Segregated Witness \cite{lombrozo2015segregated} to increase the block creation rate or increase the block size. The original data is split into signature information and transaction information, and a new data structure called ``Witness'' is formed to store the signature information separately. However, this method can only reduce the capacity pressure of Blockchain and still has a significant performance gap compared with the centralized payment system. 

The second approach is to only keep the last state of different channels on a Blockchain. In this approach, small transactions are processed in different channels off-chain to avoid occupying the Blockchain capacity. A typical example of this approach is the lightning network. For each update in the channel, the lightning network requires synchronous operation and high bandwidth. To improve efficiency and resource consumption, duplex micro-payment channel protocol is proposed. It allows users to establish an off-chain channel to realize real-time payment without delay and at the same time, guarantees end-to-end security.


\subsection{Asset trading}

Delphine N. \cite{nougayrede2018towards} argued that as a result of the financial crisis triggered by the collapse of Madoff, MF Global, and Lehman Brothers, trust in financial institutions has bottomed out as a significant amount of the world's wealth has been invested in stocks and bonds issued by large western economies. End investors are increasingly calling for a more transparent, fair, and reliable asset trading mechanism. Significant changes brought by distributed ledger technology (DLT) or Blockchain are increasingly associated with asset trading. From the issuance of securities to the settlement of transactions, Blockchain is considered more suitable in such distributed environments. Blockchain can merge these silos of information in the traditional business model into a master record.

In \cite{chiu2019Blockchain}, Jonathan C. and Thorsten V. have stated that ``Many practitioners believe that a Blockchain, or Distributed Ledger Technology (DLT), has the potential to radically transform securities settlement''. By applying Blockchain to security trading, settlements can be performed faster by getting rid of the fragmented post-trade infrastructure and allowing for more flexible settlement cycles. The authors also showed that optimizing block size and block time can guarantee the efficiency of transaction settlement. In addition, the generation of sufficient transaction fees must be made to ensure that the Blockchain is tamper-proof. If the investor wishes to settle his/her transactions earlier, larger fees should be paid to get into blocks faster. In \cite{8855688}, Lei et al. discussed applying Blockchain to a wider range of heterogeneous, non-financial digital assets (such as reward points for hotels and airlines). Lei et al. also proposed a decentralized equilibrium trading mechanism that supports three key features: decentralized general ledger, equilibrium pricing, and asset circulation. The proposed trading mechanism allows investors to trade digital assets across multiple value systems on a single platform.

\subsection{Loan}

Banks operate on large, complex centralized information systems and databases. In such an environment, the scarcity and monotony of data can severely limit the assessment of individual credit conditions. Such limitation often leads to high individual customer loan service costs, a problem that is especially acute with loans to help the poor in marginal areas. In \cite{WANG2019648}, Hao et al. proposed a loan management system based on permissioned Blockchain Hyperledger Fabric, which supports smart contracts to offer efficient information exchange, and customer data protection and prevents data tampering. In terms of personal loans, an increasing number of traditional banks are also facing competition from technology-based lenders. 

After the 2008 financial crisis, P2P platforms become substitutes for Banks. To compete with P2P companies, traditional banks need to solve unitary banking and asymmetric information problems. Besides, banks may also need to use data from online shopping platforms, tax and judicial systems, recruitment platforms, car rental companies, financial institutions, and other organizations. 

In the context of personal loan service, instead of tying up loan applicants' financial and asset status through traditional methods, such as interviews and on-the-spot visits, a Blockchain-based loan over-prevention mechanism was proposed in \cite{8949077}. Through this mechanism, the problem of sharing private customer loan information data among institutions can be solved. For corporate loans, Arantes et al. \cite{8726785} proposed the adoption of permissioned Ethernet Blockchain networks, which can improve the transparency of public fund allocation through lending, monitoring, and evaluation processes.

\subsection{Remittance}

For the past 45 years, the Society for Worldwide Interbank Financial Telecommunications (SWIFT) has served as a centralized cross-border remittance network, supporting more than 10,000 institutions to process 24 million remittance transactions per day. Compared with traditional remittance networks such as SWIFT, the Blockchain-based remittance network is decentralized, whereby financial intermediaries can be bypassed. By cutting intermediate banks' costs, remittance fee is much cheaper, and transfers can be completed in minutes or even faster. 

Ripple is a P2P network based on distributed ledger technology and cryptocurrency. Unlike the SWIFT network, the Ripple network transmits transaction information and simultaneously settles payments within seconds. Remittance through the Ripple network does not require multiple currency conversions and several days of processing by intermediaries, which greatly shortens the processing time. Remittance costs are also much lower through the Ripple network \cite{qiu2019ripple}. Ricci P. and Mammanco V. \cite{8969668} proposed RemBit, which is a Blockchain-based remittance system launched in Ethiopia. RemBit is able to transfer money while complying with Ethiopian government regulations. RemBit allows counterparties to share remittance data.

\subsection{Internet of Things (IoT) for the finance industry}

The Internet of things (IoT) connects objects in the real world through the Internet. It is an extension of things and places, changing consumers' lives. Nowadays, banks are more inclined to adopt IoT technology to enrich and diversify their financial services scenarios. Recent finance-related digital trends using IoT technologies include electronic combination locks for treasury or ATMs, biological recognition devices, electronic signature screens, collaterals and assets, real-time tracking devices, wallet of things, etc. Financial institutions use these IoT devices for security management, customer identification, authentication, personalized services, asset evaluation, risk control, and efficiency improvement.

Blockchain is a natural carrier for IoT devices. Blockchain's distributed ledger and consensus mechanisms allow two devices to communicate and exchange data point-to-point and ensure that the IoT device records cannot be tampered with. Each device in IoT uses a unique ID that can be easily tracked. However, Blockchain-based IoT systems face user identity and transaction information privacy protection risks such as de-anonymization analysis, wallet privacy disclosure, Sybil attacks, message spoofing, linking attacks, and other privacy incursions. Kosba et al. \cite{7546538} proposed ``ShadowEth'', a trusted execution environment based on the hardware-integrated Blockchain to store and execute smart private contracts. ShadowEth protects transaction privacy of smart-contract-integrated, Blockchain-based, and IoT financial applications.

\subsection{Financial inclusion}

Financial inclusion refers to providing appropriate and effective financial services at an affordable cost to all social strata and groups in need. These financial services are based on the principle of equal opportunities and business sustainability. In developed countries or regions with a more mature financial system, people generally have access to convenient financial services. However, in developing economies, more than two billion people have limited access to formal financial services. Blockchain-based FinTech can serve low-income people ignored by the formal financial system at a lower cost as well as in a more convenient and flexible way. For example, Blockchain can be used to store biometric data collected through smartphones and establish trusted identities for people and companies. These authenticated identity data can be considered as Know Your Customer (KYC) credentials. Based on this identity information, people can participate in financial inclusion services without face-to-face verification.

\begin{figure}[hbt]
  \centering
  \includegraphics[width=0.45\textwidth]{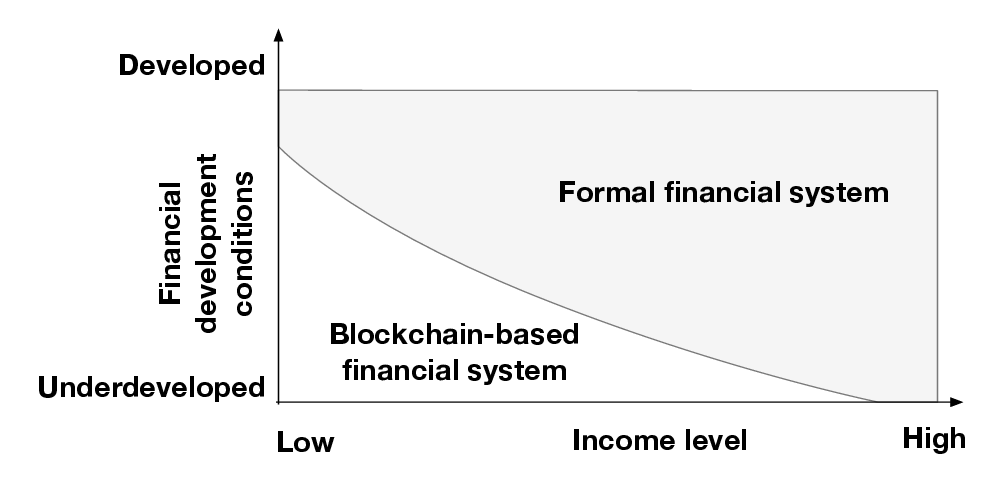}
  \caption{Financial inclusion preference under different financial development conditions and income levels.}
  \label{InclusiveFinance}
\end{figure}

In \cite{larios2017Blockchain}, Guillermo et al.  identify several financial inclusion sensitivities such as cash preferences, lending practices, transfers and remittances, identification, and incumbent system limitations. The authors suggest that financial inclusion is closely related to financial development conditions and income level. If the financial infrastructure is complete and the income level of individuals is improved, they will usually choose formal financial services. In contrast, the preferences for informal financial practices are growing due to the lack of bank branches, poor automatic teller machine (ATM) coverage, and low income. As depicted in \reffig{InclusiveFinance}, Blockchain has wider application scenarios in areas with low-income groups and underdeveloped financial development conditions. 

Sebastian et al. \cite{schuetz2019Blockchain} argued that geographical access, high costs, inappropriate banking products, and financial ignorance are the four main challenges hindering financial inclusion. To overcome some of these challenges and to alleviate rural financial exclusion in India, Sebastian et al. analyzed the factors that drive and inhibit the adoption of Blockchain solutions. Factors include usefulness and usability, user and provider characteristics, environment variables, interrelationships of adoption, degree of adoption, etc. Firdaus et al. \cite{8969847} also introduced Hyperledger Fabric financial record system to manage savings and loans for cooperatives in Indonesia. This system is designed to prevent embezzlement of funds by taking advantage of the distributed ledger properties of Blockchains to increase the transparency of transactions.

\subsection{Enterprise-level interaction in finance}

Many areas of financial services involve an extensive degree of paper-based processing and manual inspections. Inter-organizational processes often have numerous intermediaries. Therefore, these organizations are facing the problems of low efficiency, regulation risks, and high costs. By applying Blockchain-based smart contracts, procedures that rely heavily on paperwork can be digitized and automated. Such applications can also reduce manual intervention and improve processing efficiency.

In \cite{weber2016untrusted}, Weber et al. proposed an approach to compiling the flow control and business logic process model into smart contracts to ensure that the entire process is executed according to established rules. The so-called trigger component allows these inter-organizational process implementations to interact with external systems or internal process implementations through Web services. In \cite{garcia2017optimized}, García-Bañuelos et al. proposed an optimized solution that translates a Business Process Modeling Notation (BPMN) process model into a minimized Petri net, then compiles this Petri net into a smart contract. The aim of this approach is to reduce deployment and execution costs for executing collaborative processes. The Blockchain-based business process management system reported in \cite{lopez2019caterpillar} was also designed to support subprocesses, boundary events, multi-instance activities, and several types of BPMN tasks.

In the context of supply-chain finance, the application of Blockchain technology in inter-organizational process management could greatly reduce manual intervention. With suppliers, buyers, and banks as the main stakeholders, they could share contractual information on a decentralized distributed ledger. Adopting smart contracts could also ensure that payments are made automatically once the contract terms are met. Increased transaction efficiency can bring additional benefits to every participant in the supply chain.

\subsection{Governance}
Unlike the Internet, Blockchain does not have international organizations that regulate its behavior. For financial service areas, which take stability and trust as the cornerstone, lack of regulation limits the application of Blockchain technology. While Blockchain is one of the most decentralized underlying technologies ever developed, it is governed by the votes of its communities. Therefore, Blockchains are more likely to be influenced by a few major contributors to the community. 

Recently, ``R3'' launched ``Corda'' which is more analogous to a ``distributed ledger technology'' (DLT) rather than a Blockchain. ``R3'' is composed of a number of leading global financial services institutions, technology companies, central banks, regulators, and trade association members. ``R3'' is a step towards a system with better governance. So far, Blockchain governance represented by Ethereum is realized through technical code, which can cause inflexibility, insecurity, and escalation of risk. In this regard, Blockchain governance represented by Ethereum does not provide financial institutions with enough confidence in the large-scale adoption of Blockchain in their businesses. Due to this weak governance mechanism, Blockchain cannot completely replace the current payment system or realize the dis-intermediation of financial institutions at this point. However, these obstacles are not likely to hinder the attempts by financial institutions to integrate Blockchain into their existing businesses.

\section{Obstacles to widespread adoption of Blockchain}

\textcolor{black}{With the development of Internet technology, the demand for different types of financial products and new services has significantly increased. Under the intense competition, financial institutions must strive to reduce operating costs while providing an appropriate level of personalized services. To this end, financial institutions must be equipped with the relevant technology to develop mass-customizable products or services.}

\begin{figure}[htbp]
  \centering
\includegraphics[width=0.4\textwidth]{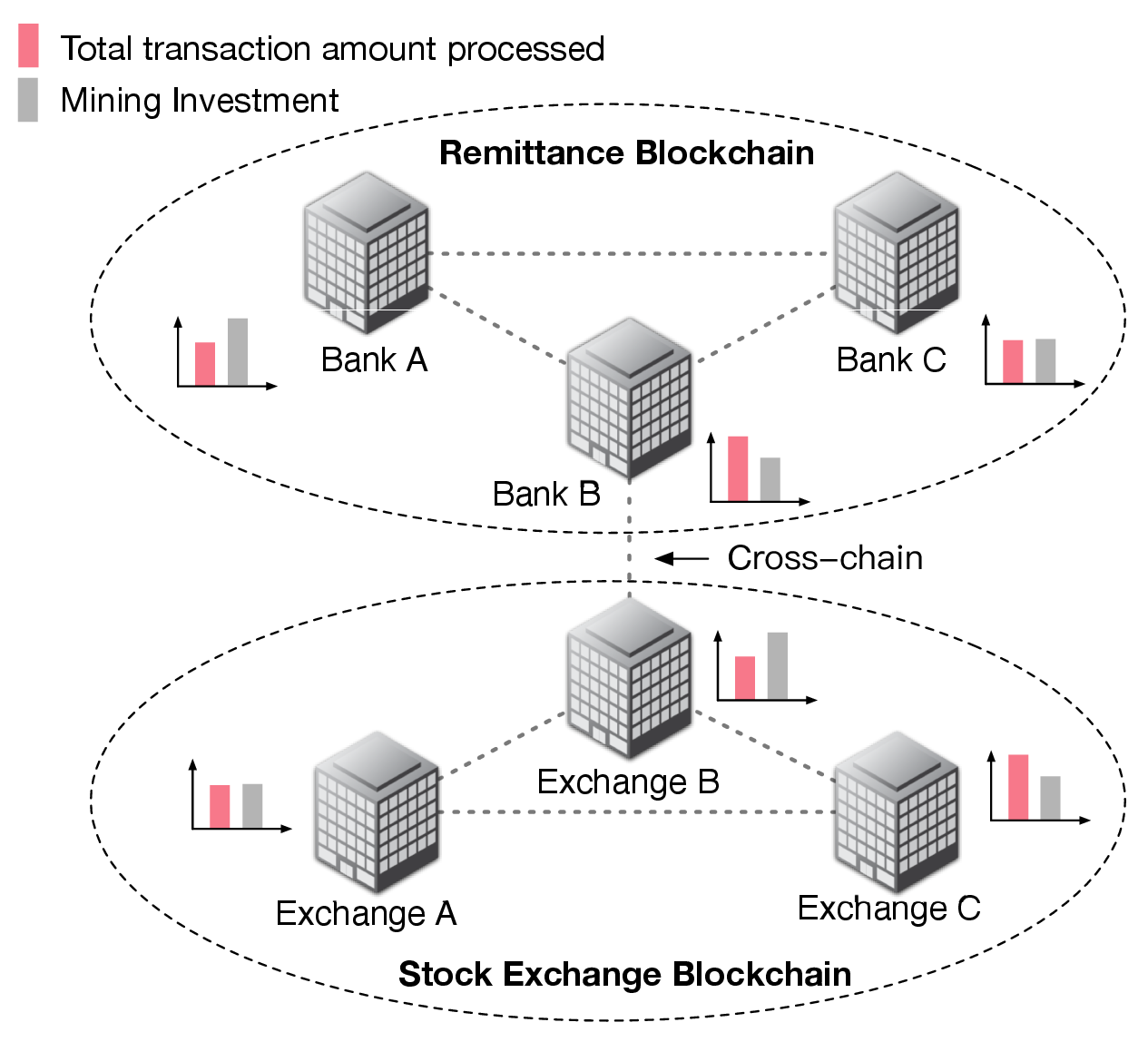}
\caption{Schematic diagram of the conflict between remittance and stock exchange when adopting Blockchain.}
\label{fig1}
\end{figure}

The need for mass customization drives financial institutions to adopt strategies that can improve efficiency. Blockchain-based applications in financial service areas can improve operational efficiencies and synergies. This means that no single company can control the costs on its own and that many financial institutions can work together to reduce costs for the industry as a whole. However, there are many obstacles preventing Blockchain from being widely adopted in financial service areas. \reffig{fig1} depicts the Blockchain networks in two different financial services areas: banking and securities trading. We analyze the different scenarios in \reffig{fig1} to help us better understand the incentive-related obstacles in detail.

As shown in the upper half of \reffig{fig1}, three banks -- Bank A, Bank B, and Bank C -- form a Blockchain network to process remittance transactions. The transaction revenue generated on the Blockchain still follows the revenue model of the original business, and banks like Bank B with large-amount transactions can still receive higher transaction fee revenue. As Blockchain nodes, even Bank A, with small-amount transactions participating in the Blockchain network, may invest more hardware resources to win more mining competition. However, because Bank A's transaction amount is smaller than those of other banks, Bank A still does not receive a proportionate amount of investment revenue from its involvement in mining activities.

Blockchain technology has long been considered an underlying technology that can connect with various financial service scenarios. Due to the conflict of revenue models, stakeholders cannot form an effective incentive mechanism to participate in the construction of the Blockchain network jointly. \textcolor{black}{Although there exist consortium Blockchain products such as R3, Libra, UBS-led Utility Settlement Coin, J.P. Morgan's Digital Coin for Payments, etc. financial institutions prefer to develop their own Blockchain standards to increase their influence and dominate the market share.} As shown in \reffig{fig1}, the remittance Blockchain may adopt different Blockchain technologies than the stock exchange network, leading to cross-chain problems. As a result, there are many obstacles to forming a stable and widespread Blockchain network that can host multiple banking services to support asset digitization and peer-to-peer value transfer, thereby reconstructing the financial infrastructure.

\section{Threats of using transaction fees as mining rewards}

The security of Blockchain's consensus protocol relies on miners behaving correctly, and miners are incentivized to do so via mining incentives. For public Blockchain, the mining rewards consist of two parts, block rewards and transaction fees. However, existing financial institutions favored permissioned Blockchain, in which trusted parties have been designated to update and manage stored information. Since permissioned Blockchain can not rely on tokenized money balances issued by a central bank like a public Blockchain, transaction fees become the sole source of mining rewards.

Next, under what conditions is it feasible to conduct transactions on the Blockchain that acts as a financial services infrastructure? In banking and finance, large-amount transactions usually require a higher settlement priority. Whereas under Blockchain, to control the speed of settlement, participants will select how fast they would like to settle by posting transaction fees to have their transactions incorporated into a block. Therefore, a proportional fee is a more suitable option rather than submitting a fixed per-transaction fee for mining incentives.

Financial institutions can set aside a portion of the transaction fee as mining rewards to discourage incentives to fork the chain. By splitting transaction fees, financial institutions maintain not only a profitable position in the volume of business but also get their fair share of mining rewards according to computing resource investments. The transaction fee pledged by investors $F$ is composed of two parts: (a) the fee charged under the existing model, denoted by $F_t$, and (b) the fee contributed as mining incentives denoted by $F_m$. The transaction fee distribution ratio is denoted as $p$.  

\vspace{-0.3cm}
\begin{equation}
F=
\begin{cases}
F_t=F*p& \text{ Fee under existing model } \\
F_m=F*(1-p)& \text{ Fee as mining incentives }
\end{cases}
\end{equation}

\linespread{1.15}
\begin{table*}[hbp]
  \footnotesize
  \centering
  \caption{Blockchain comparison and their relevance to Asset trading.}
  \label{tbl1}
    \begin{tabular}{p{2.8cm} p{3.2cm} p{0.8cm} l p{1.2cm} p{2cm} p{0.8cm} p{1.2cm}}
      \hline
      Platform & Network Type & Support NFT & Crypto-currency & Smart Contract & Consensus & Mining Cost & Throughput \\
      \hline\noalign{\smallskip}
      Bitcoin & public, private, consortium & no & built-in (Bitcoin) & no & PoW & high & low \\
      \specialrule{0.00em}{1pt}{1pt} 
      Ethereum & public, private, consortium & yes &  built-in (Ether) & yes & PoS & low & Good \\
      \specialrule{0.00em}{1pt}{1pt} 
      Litecoin & public, private, consortium & yes & built-in (Litecoin) & yes & PoW & low & Good \\
      \specialrule{0.00em}{1pt}{1pt} 
      EOS & public, private, consortium & yes &  built-in (EOS) & yes & DPoS & low & Excellent \\
      \specialrule{0.00em}{1pt}{1pt} 
      Stellar & public, private, consortium & yes & built-in (Lumen) & yes & stellar consensus & low & Good \\
      \specialrule{0.00em}{1pt}{1pt} 
      R3Corda & private, consortium & yes & not built-in & yes & BFT & no & Good \\
      \specialrule{0.00em}{1pt}{1pt} 
      Quorum & private, consortium & no & not built-in & no & majority voting & no & Good \\
      \specialrule{0.00em}{1pt}{1pt} 
      MultiChain & private, consortium & no & not built-in & no & PBFT & no & Good \\
      \specialrule{0.00em}{1pt}{1pt}  
      Hyperledger Fabric & private, consortium & yes & not built-in & yes & PBFT & no & Good \\
      \noalign{\smallskip}\hline
    \end{tabular}
\end{table*}

However, when transaction fee dominates mining rewards, strategic deviations such as \textsl{Selfish Mining}, \textsl{Undercutting}, \textsl{Mining Gap}, \textsl{Pool Hopping} could threaten the integrity and security of the Blockchain \cite{10.1145/2976749.2978408}. In response to the threats of transaction fees becoming a growing share of mining rewards, EIP1559 \cite{EIP1559} (also known as ``London Hard Fork") was launched on Ethereum on August 5, 2021. EIP1559 includes a base fee representing the minimum gas price a user needs to pay in each block. EIP1559 dynamically adapts and burns based on the parent block to reduce the impact of transaction fees on block incentives. EIP1559 allows users to specify the transaction fee bids. Users can set a priority fee per gas to incentivize miners to prioritize their transactions. One of the characteristics of EIP1559 is that the base fee is always burned, and the miner can only keep the priority fee. Therefore, it removes the incentive for the miners to manipulate the fee. 

In Ethereum, the value of these burned ETH is transferred to the scarcity of existing ETH, causing ETH deflation. However, this approach is obviously not applicable in the case of Blockchain as a financial infrastructure. The transaction fee originally generated through economic activities cannot vanish through the burning process but transfer to a central organization, which will obviously have an adverse impact on the development of Blockchain. In the scenario of Blockchain as financial infrastructure, another more suitable solution called Dynamic Transaction Storage (DTS) \cite{9592512} strategy is proposed. DTS strategy dynamically adjusts the number of transactions incorporated into a block base on each transaction's fee amount. As a result, the DTS strategy can maintain a stable ming reward for different blocks and removes the incentive for the miners to manipulate the fee.

As for now, a Blockchain can generate fees by limiting the speed at which it is updated. The traditional approach is through the design of the Blockchain by restricting the block size (how many transactions can be included in each new record) and the block time (how frequently new records are incorporated). Whereas under the DTS strategy, by incorporating high-fee transactions into smaller blocks, investors will achieve a low transaction consensus latency that matches the transaction fee pledged. To makes fast settlement a scarce resource, DTS no longer relies on restricting the block size and the block time to limit Blockchain throughput but rather on the level of network propagate time required to reach consensus. This lays a good foundation for improving Blockchain scalability by adjusting block size and block time.

\section{Blockchain selection for financial services }

Blockchain technology adoption in financial services could mitigate risks and encourage competition, consumer protection, financial integrity, and financial stability. In Table~\ref{tbl1}, we summarize the classification of Blockchain platforms and compare them from different perspectives, including network type, support of NFT, support of cryptocurrency, enabling smart contracts, consensus type, mining costs, and throughput. Through the comparison, we hope to identify the most suitable Blockchain platform for the financial sector.

\begin{itemize}

\item \textbf{\textsl{Network Type}} For a permissioned Blockchain, only trusted validators can update the Blockchain. However, validators still need to behave honestly, which can be achieved through economic incentives. For a permissionless Blockchain, the consensus protocol and the reward scheme must also be designed properly to prevent users from tampering with the Blockchain. The comparison between the two systems thus depends on the degree of commitment and enforcement. Hence, whether DLT can reap significant advantages relative to existing, centralized financial service systems will depend on the costs of providing such incentives versus designing a tamper-proof consensus protocol.

\item \textbf{\textsl{Support NFT}} Financial services such as asset trading, which is built around distinct units, could benefit from units existing on a Blockchain as NFTs. NFT protocols are constantly evolving, ERC721 (Ethereum Request for Comments 721) \cite{EIP721} helps us find a way to present distinctive details about an asset in the form of a token. Whereas the ERC1155 (Ethereum Request for Comments 1155) \cite{EIP1155} standard outlines a smart contract interface that can represent any number of fungible and non-fungible token types. It is more suitable to represent different kinds of financial assets with a certain quantity, such as securities, insurance, fund, etc.

\item \textbf{\textsl{Smart Contract Enabled}} Smart contracts are self-executing programmable contracts that encode an agreement between two or more parties. The terms of a transaction are written as a protocol that exists across a distributed, decentralized Blockchain network system. They are also changing the face of the banking industry in the processing of insurance claims, stock trading delivery vs payment, transparent auditing, and so on.  

\item \textbf{\textsl{Consensus}} Within financial services, Proof-of-Work (PoW), Proof-of-Stake (PoS), and Delegated PoS (DPoS) are some of the more popular consensus mechanisms in public Blockchains, while Byzantine Fault Tolerance (BFT) \cite{gramoli2020blockchain}, Practical Byzantine Fault Tolerance (pBFT) \cite{castro1999practical}, and LibraBFT \cite{baudet2019state} are popular in private Blockchains. PoW is a secure and resilient method of forming consensus, but it consumes significant energy and can be slow and expensive during times of high network traffic. PoS improves the weaknesses of large energy consumption in PoW consensus mechanisms while preserving network security. BFT, pBFT, and LibraBFT offer an instant settlement. However, they raise new regulatory concerns in areas such as competition that run counter to the core "ideal" of decentralized networks.

\item \textbf{\textsl{Mining Cost}} The increasing use of Blockchain in financial services aims to bring benefits to many stakeholders. Whereas the use of energy-intensive methods to achieve consensus, such as PoW presents unacceptable risks to financial stability and society. Therefore consensus protocols such as PoS, DPoS, and BFT are more practical consensus choices to serve the financial sector effectively.

\item \textbf{\textsl{Throughput}}  In the financial field, transactions are continuously carried out around the account level, if the transaction processing speed is too slow, it may cause inconsistent accounting processing. For example, a selling transaction depends on the account balance, but the balance depends on the settlement of the previous buying transaction. Inconsistencies can lead to risks to settlement finality. Transactions between counterparties carry risks—including credit, liquidity, operational, and legal risks—all of which can trigger systemic hazards.  Greater transaction throughput is crucial for Blockchain adoption in financial services contexts.

\end{itemize}

From the comparisons listed in Table ~\ref{tbl1}, we conclude that EOS \cite{ethereum.org} is the superior Blockchain platform of choice to support financial and banking transactions with the goal of improving performance and reliability. Although the private and consortium Blockchain has been favored by existing financial intermediaries, public Blockchain is better suited for an integrated, secure global marketplace for the exchange of different financial assets. Also, EOS benefits from DPoS consensus include energy savings, greater decentralization, high throughput, and positive participant behavior. DPoS facilitates greater democratization, voting is proportional to the number of shares cast by participants. DPoS also aims to promote positive behavior among participants. Well-behaved participants are more likely to be voted repeatedly. Any witness who loses credibility, does not participate, or appears to engage in fraudulent activity, is likely to be voted out.

Nevertheless, as a financial services infrastructure, the rewards that underpin its operations should come from the fees generated from the transactions processed on it. Due to the volatility of fees incurred by various transactions, the DTS strategy should be implemented on top of the EOS Blockchain platform in order to form a more equitable block reward mechanism. We believe this is a superior way to get more financial services running on DLT and leverage the benefits of DLT while ensuring appropriate risk mitigation.

\section{Conclusion}

In order to support the rapid development of financial services in the Greater Bay Area and the complex business environment, Blockchain has the potential to disintermediate markets, increase efficiency and speed, and reduce costs. In this paper, we describe a comprehensive survey about the current application status of Blockchain in financial service areas. This paper covers key areas in the financial industry, such as insurance, banking, payments, asset trading, loans, remittances, the Internet of Things (IoT) for the finance industry, financial inclusions, enterprise-level interaction in finance, and governance. From our analysis, one can see various ongoing efforts in integrating Blockchain technology into traditional and emerging financial service areas.

When different financial institutions participate in the establishment of the financial Blockchain, they are faced with a realistic problem of revenue distribution. We believe that it is the most appropriate choice to use transaction fees generated from economic activities on the Blockchain infrastructure as mining rewards. However, miners have the incentive to deviate from current norms to maximize their rewards from mining when transaction fees dominate mining rewards. We analyzed EIP1559 and DTS strategy and concluded that the DTS strategy is a better solution because it can distribute profits more fairly and avoid deviant mining behavior.

Finally, we compared the current mainstream Blockchain through network type, support NFT, mining cost, throughput, and other aspects. We concluded that Blockchain with features like EOS and integrated DTS strategy is the more promising option for the financial Blockchain infrastructure in the Greater Bay Area.

\vspace{-0.2cm}
\section*{Funding}
This research was funded by the University of Macau (file no. MYRG2022-00162-FST).

%


\end{document}